# An Efficient Transport Protocol for delivery of Multimedia Content in Wireless Grids

Srinivasan Arulanadam[1], Suresh Jaganathan[2]*, Damodaram Avula[3]
1. Department of Information Technology, Misrimal Navajee Munoth Jain Engineering College, Anna University, Chennai, Tamilnadu, India, Email: asrini30@gmail.com
2. Department of Computer Science and Engineering, Sri Sivasubramania Nadar College of Engineering, Anna University, Chennai, Tamilnadu, India
3. Department of Computer Science and Engineering, Jawaharlal Nehru Technological University, Hyderabad, Andra Pradesh, India, Email: damodarama@jntuh.ac.in
* E-mail of the corresponding author:   whosuresh@gmail.com

**Abstract**
A grid computing system is designed for solving complicated scientific and commercial problems effectively, whereas mobile computing is a traditional distributed system having computing capability with mobility and adopting wireless communications. Media and Entertainment fields can take advantage from both paradigms by applying its usage in gaming applications and multimedia data management. Multimedia data has to be stored and retrieved in an efficient and effective manner to put it in use. In this paper, we proposed an application layer protocol for delivery of multimedia data in wireless girds i.e. multimedia grid protocol (MMGP). To make streaming efficient a new video compression algorithm called dWave is designed and embedded in the proposed protocol.  This protocol will provide faster, reliable access and render an imperceptible QoS in delivering multimedia in wireless grid environment and tackles the challenging issues such as i) intermittent connectivity, ii) device heterogeneity, iii) weak security and iv) device mobility.

## 1. Introduction

Grid technology (Dietmar Erwin, 2001) (Foster, I., 2001) is a new paradigm which has the potential to completely change the way of computing and data access. Computational Grids are widely regarded as the next logical step towards computing infrastructure, tightly linked clusters, enterprise-wide clusters, and geographically dispersed computing environments following a path from standalone systems. Generally speaking, the grid could be considered as a new enabling technology to transparently access computing and storage resources anywhere, anytime and with guaranteed quality of service. Grid is already being successfully used in many scientific applications to store and process the vast amount of data. Examples include the processing of data coming from physic-related applications such as the data generated in the nuclear accelerator of CERN (CERN, 2008), the data coming from radio telescopes, the data generated in complex simulations (Datagrid, 2004), and so on.

Applications using multimedia streaming over wireless networks have been increasingly deployed and available due to the advent of 3G and 4G type communications. However, delivery of multimedia techniques are designed for wired networks and perform poorly over wireless networks. Our research focuses on the impact of a using grid computing in wireless and to improve the performance of streaming multimedia applications. In order to adopt the requested multimedia service, video-on-demand (VOD)  systems (L. M. Nithya and A.Shanmugam, 2010) have two types of tailoring techniques, i) the client-side adaptation and ii) the server-side adaptation. Client-side adaptation suffers from resource overload restriction, and unnecessary channel overload due to the transmission of high-quality media. On the other hand, a scalability issue due to the large number of users affects the server-side adaptation. Grid computing represents a powerful way to overcome these problems, due to its distributed nature and is possible to create robust and scalable systems without overloading user devices.

VOD system requires enormous storage capacity, and a centralized solution which has several disadvantages, including high infrastructure costs (L. M. Nithya and A.Shanmugam, 2010). For these reasons, in recent years the efforts of researchers have been shifted towards improving multimedia server performance by designing and developing radically with different solutions. The grid computing seems to be a good platform for implementing such a kind of systems as it supports scalability, robust, reliability and has in-build security. A grid-based multimedia streaming and service's protocol in an application layer is proposed in this paper. This protocol allows to reduce the provisioning delay experienced by users and also to optimize the resource utilization by reducing the idle time of the computational nodes engaged in the distributed tailoring (Bruneo D et.al, 2009).

This paper aims at showing a real experience on merging grid computing and wireless environment with grid computing as an underlying technology, in order to promote a new computing paradigm to deliver multimedia and its services in wireless networks. The paper is divided into 6 sections. Section 2 does a detailed study on wireless girds. Section 3, briefs basic concepts on protocols used in wireless and grid networks. Section 4, describes the architecture of the proposed application layer protocol. Section 5 reveals experimental results and does performance analysis, and Section 6 concludes the paper.

## 2. Wireless Grids

Grid computing continues to attract a large amount of attention in many applications. As it shares the resources among various geographically located participants, the concept of grid has gradually been extended to the wireless world. A wireless grid is a modified form of wired grid to facilitate the exchange of information and interaction between heterogeneous wireless devices. It comes with added complexities of wireless devices such as limited bandwidth, dynamic in nature, and the need for quality of service. Depending on the method of interaction in wireless grid, various layouts can be envisaged. The functionality and ability of these methods in a wireless grid are contingent upon the resolution of many technical challenges of the grid. The possibility of wireless grid has to meet some criteria, such as needs at the client side and various service levels. These criteria further fueled by technological advances in two areas, i) Grid Computing and ii) Wireless Communications

**Grid Computing**: It is defined as "flexible, secure, and coordinated resource sharing among dynamic collections of individuals, institutions and resources." The main goal of the grid is to provide i) secure, ii) inexpensive, and iii) real-time access to dynamic, heterogeneous resources, potentially around geographic boundaries. The technologies originally developed for use in a wired environment and are now being augmented to operate in wireless situations. And in wireless situations, it has to maintain the desirable characteristics such as stability, transparency, scalability and flexibility.

**Wireless Communications**: Rapid advances, diminishing prices, wide availability, and attractive form factors have caused wireless technologies to permeate the lives of people from all walks of life. The development of the wireless technologies such as 802.11, GPRS and 3G has extended the reach of wireless services to (literally) the man in the street. With the ubiquity and indispensability of wireless technologies established, these technologies are now making inroads into grids. One way to characterize the architecture of wireless grid is by finding the degree of heterogeneity and level of control exercised in the devices. It can vary from a simple network of homogeneous devices bound by a single set of policies and rules to a complex network of heterogeneous devices spread across multiple organizational and geographical boundaries. The architecture of wireless grid can be categorized into two types, i) Local Cluster or Homogeneous and ii) Wireless Intra-Grids.

**Local Cluster or Homogeneous Wireless Grid:** Figure 1 shows the local cluster or homogeneous wireless grid. It involves, with its simplest form, a local collection of identical or similar wireless devices that share the same hardware architecture and the same operating systems. It is because of the homogeneity of the end systems, the integration of these devices into the wireless grid as well as the consequent sharing of resources becomes a much easier task. Today, this type of organization is more likely to be found in a single division of an organization where one single administrative body exercises control over all the devices. An example would be a network of mobile handheld devices for coordinating position of soldiers in an army field.

**Wireless Intra-Grids:** Figure 2 shows the wireless Intra-Grids cluster architecture. An intra-grid consists of wireless devices that belong to multiple divisions or groups within an actual organization (AO). The divisions may be located in different geographies and have its own set of policies, level of trust and oversight so that ground truth may be known with respect to identity and characteristics. AOs are the point where resolution can occur between the virtual presence of a wireless entity and its actual name and location. AOs also tend to be persistent in time, and become the point of composition among other AOs. An example of an intra-grid would be a wireless grid that simultaneously supports the mobile sales force of a company and tracks the inventory using networks of wireless sensors for the manufacturing division.

## 3. Protocols used in wireless and grid environments

Many wireless network applications require efficient delivery of multimedia content capabilities. However, the existing transport protocols mainly aim to achieve high throughput, efficiency and reliability objectives only. These protocols do not address multimedia communication challenges in a wireless network. In a computational grid, the data to be processed can be a multimedia data. Those data should be transferred to job nodes for processing. Available in-use grid protocols are good at transferring texts. For delivery of multimedia protocols such as RTP, RTCP, RTSP and Media Streaming protocol (MSP) (Christopher K. Hess and Roy H. Campbell, 1999) support to some extent and lacks in speedy transfer. In (Suresh Jaganathan, Srinivasan A, A Damodaram, 2011) authors presented a performance comparison for different transport protocols that can be used in grid computing. As our work involves in developing efficient transport protocol for delivery of multimedia in wireless grids, we used three protocols i.e. JXTA, GridFTP and UDT.

*3.1. JXTA*
JXTA is an open source P2P technology introduced by Sun Microsystems. It enables developers to create different kinds of services and applications and is a kind of infrastructure that hides the complexity of underlying networks and making all kinds of devices interoperable. It has three design principles, i) interoperability, ii) platform independence and iii) ubiquity. It has a set of open, generalized peer-to-peer protocols, which allows any networked devices such as PDAs, laptops, workstations, servers and supercomputers to communicate and collaborate mutually as peers (Emir Halepovic and Ralph Deters, 2003). JXTA has a set of protocols defined by simple XML messages. The fundamental concepts for the protocol are

*Peer* - end node or end point or a participant with an unique ID
*Peer Group* - a collection of end nodes with its own policy
*Advertisement* - this contains the description about the existing resource
*Message* - communication between two end points
*Pipe* - communication mechanism to transfer message between two end points.

*3.2. GridFTP*
GridFTP has been used as a data transfer protocol for effectively transferring a large volume of data in grid computing (John Bresnahan et.al, 2007). It supports a feature called parallel data transfer that improves throughput by establishing multiple TCP connections in parallel and automatic negotiation of TCP socket buffer size. However, to achieve high throughput, the number of TCP connections should be optimized according to network condition. In order to get a maximum data transfer throughput, it has to use optimal TCP send and receive socket buffer sizes for the link being used. The optimal buffer size is twice the bandwidth delay product of the link defined as Buffersize = 2*bandwidth_delay. The Globus (Rajeev Wankar, 2008) with GridFTP provides a software suite optimized for the gamut of data access issues from bulk file transfer to the details of getting the data out of complex storage systems within sites on the grid.

*3.3. UDT*
UDT is an application level, end-to-end, unicast, reliable connection oriented data transport protocol (Yunhong Gu & Robert L. Grossman, 2007). The UDT protocol is completely at user space above UDP. It uses UDP for transferring user data and TCP for control information. UDT uses packet based sequencing to check packet loss and guarantee data reliability. It is specially designed for high speed bulk data transfer in order to remove and reduce the overhead of memory copy, loss information processing, acknowledging, etc. UDT has a congestion control mechanism which maintains efficiency, fairness, and stability. And it can be deployed in low cost, as it is in the application layer. packet based sequencing, and selective positive acknowledgement (ACK) is sent at every constant interval, whereas negative acknowledgement (NAK) is generated as soon as packet loss is detected.

4. **Architecture of the proposed protocol**
Figure 3 shows the architecture of proposed application layer protocol for multimedia streaming and services in wireless grids. Architecture consists of four layers, i) layer 1 constitutes the fundamental elements of communication needs for the network, ii) layer 2 contains grid architecture and uses grid middleware called globus, which form a communication layer for layer 3 and upper layers and take care of grid oriented services such as data storage and security, iii) layer 3 consists of open source, virtual protocol for communication between peers who can act as both server and client. It uses a middleware called JXTA, which forms the basic communication between wireless peers, iv) layer 4 contains the proposed protocol which combines wireless and grid, thus enabling grid environment in wireless networks. This layer contains sub-layers, a) a new compression module for compressing multimedia files (dWave), b) security module, which helps to stream securely and c) QoS module, providing quality of service in streaming multimedia data, which checks whether the multimedia has viewable quality, low jitter, high throughput. In layer 5, applications can be developed by using the underlying four layers for delivery of multimedia streaming and its services in wireless grids. Forthcoming paragraphs give detailed description of each layer.

*4.1. Wireless Middleware Layer*

JXTA technology is a network programming and computing platform (Esteve Riasol & F. Xhafa, 2006) designed to solve a number of problems in modern distributed computing, especially in peer-to-peer computing. JXTA is originally conceived and designed with the participation of a growing number of experts from academic institutions and industry. JXTA defines a common set of protocols for building wireless peer to peer applications to address the recurrent problem with existing wireless networking systems of creating incompatible protocols.

It is an open network computing platform, providing building blocks and services required to enable anything anywhere application connectivity. Its building blocks help in building applications in client-server or web based computing or in distributed computing models. It has a common set of open protocols backed with open source reference implementations for developing peer-to-peer applications and are standardized in the manner, in which peers can: i) discover each other, ii) self-organize into peer groups, iii) advertise and discover network resources, iv) communicate with each other and v) monitor each other.

These protocols are independent of programming languages, and transport protocols alike. The protocols can be implemented using programming languages such as i) Java, ii) C/C++, iii) .NET and iv) Ruby. Furthermore, they can be implemented on top of TCP/IP, HTTP, Bluetooth, and other network transports all the while maintaining global interoperability. Primary design principle of this middleware is to provide a platform that embodies the basic wireless network functions in peer networks. The functionality supported by this middleware are:

1) Interoperability - enable peers to locate services and communicate with one another, independent of network addressing and physical protocols
2) Platform independence- independent of programming languages, network transport protocols, and deployment platforms and
3) Ubiquity- accessible by any device with a digital heartbeat

And it empowers end points (peers/nodes) on the edge of the network by providing a unique addressing scheme (ID). The nodes with unique IDs can migrate across physical networks, changing transports and network address, even being temporarily disconnected, and still be addressable by other end points (Gabriel Antoniu, et.al., 2005).

*4.2. Grid Middleware Layer*

Grid computing has been an active research area for many years, and a number of systems exist that utilize functional computational grids. The most notable of this is the NASA Information Power Grid (Rajeev Wankar, 2008) (run on the Globus toolkit) and the new grid being constructed for analyzing data from the Large Hadron Collider project at CERN (CERN, 2008). Computational grids has given developers a considerable number of extra problems to overcome in order to make them work correctly, reliably and also to build new middle-wares apart from Globus, which is widely used. The Globus toolkit is designed to enable people to create computational grids. Globus is an open source initiative aimed at creating new grids capable of the scale of computing. As an open source project any person can download the software, examine it, install it and hopefully improve it.

The Globus toolkit itself is made from a number of components. Figure 4 shows the three main components in the globus toolkit, i) Resource management (GRAM), ii) Information services (MDS) and iii) Data management (GridFTP/MMGP). Grid Security Infrastructure acts as a base for the three components.

*Resource management:* The resource management provides support for, i) resource allocation, ii) submitting jobs: Remotely running executable files and receiving results and iii) Managing job status and progress.

*Information services:* The information service provides support for collecting information in the grid and for querying this information, based on the Lightweight Directory Access Protocol (LDAP).

*Data management:* The data management provides support to transfer files between machines in the grid and for the management of these transfers.

The design of the toolkit itself is very modular and has been developed in a way to make alterations and improvements easier, with less impact on connected components. It is designed to work on a number of platforms, predominantly that of Linux but with limited support for Microsoft. So far, Globus has been a lead contender in the development of grid computing and is currently the only major effort with open source availability. The Toolkit itself is designed to work in research environments, predominantly as an impetus to be redesigned and improved.

*4.3. Proposed Protocol Layer (MMGP)*

The Globus middleware (Figure 5) has become the de-facto standard for Grid computing and provides the basic building blocks to do most of the things when building a grid. The components in globus middleware are i)

GRAM, ii) MDS, iii) GRIS and iv) GridFTP. First component GRAM stands for Globus Resource Allocation Manager. Its job is to manage job requests and to execute and monitor them on remote machines. GRAM has overall control of all the other services. It is responsible for setting up and taking down services provided by Globus when it believes a user needs them. The Meta Data Service (MDS), second component controls all information pertaining to the different machines on a grid. It holds information of both dynamic and static nature. Examples include machine ID, average load, memory capacity, etc. Third component, GRIS servers can be located at various points across a grid. They are designed to hold information about any machine that has been registered with them. The information in question could be either static or dynamic, and the architecture of the GRIS server is designed to be easily extendable to provide a holding space for data of any kind about individual machines. Fourth and last component, GridFTP is a protocol intended to be used in all data transfers on the grid and provides a secure and reliable data transfer among grid nodes. It is based on FTP, but extends the standard protocol with facilities such as multi-streamed transfer, auto-tuning, and Globus based security.

Figure 6 shows the place where our proposed protocol is incorporated in the globus middleware. The incorporation of our work is done with help of Globus XIO framework (Figure 7). Globus XIO is broken down into two main components, framework and drivers. The Globus XIO framework manages IO operation requests that an application makes via the user API. The framework's work is not to deliver the data in an IO operation nor manipulate of data. These works are done by the drivers. The framework's job is to manage requests and map them to the drivers interface. It is the drivers themselves that are responsible for manipulating and transporting the data. A driver is the component of Globus XIO that is responsible for manipulating and transporting the user's data. There are two types of drivers, transform and transport. Transform drivers are those that manipulate the data buffers passed it via the user API and the XIO framework. Transport drivers are those that are capable of sending the data over a wire.

*4.4. Operational block Diagram for MMGP*

Figure 8 shows the operational block diagram of proposed protocol. Consider there are two Nodes A and B. Node A starts its application by creating an input pipe which establishes endpoint or out port for Node A (1). Node A starts listening for incoming calls. Also it searches for nearest Super Node. Super node is a node which adopts a resolver policy to use a decentralized, centralized or hybrid approach to match their requirements. This policy provides the ability to send and propagate queries, and receive responses, authenticate and verify the credentials and drops invalid messages. These types of nodes do not replicate edge peer's advertisements, nor propagate queries to the edge peers. The jobs of a super node are

i)       maintaining the list of closest other super nodes,
ii)      ii) maintaining information about closest nodes,
iii)     iii) formation of route between two end points. Super Node uses hash table to maintain the details of available nodes. After finding a super node, Node A generates an advertisement which contains details about Node A and advertises it in the network (2). Super Node listens and stores the details about Node A (3) in its hash table.

Node B starts its application. It creates a search query and listens for incoming calls (4). The search query is sent to nearest Super Node (5), where a super node stores the details about Node B (6). Already super node has the details of Node A and it checks with the search query, received from the Node B, it matches and sends the query to Node A for acceptance (7). Node A gets the details about Node B and connection is established and delivery of multimedia service takes place (8).

*4.5.     Pseudo Code for Proposed Protocol [MMGP]*

This section explains the steps involved in establishing connection between two wireless grid nodes using MMGP with the help of working block diagram (Figure 9).
1) Node A creates an input pipe for a service
    a) Associate the input pipe with an endpoint or an out port.
    b) Starts listening to the endpoint.
    c) Creates an advertisement and adds its IP details.
2) Setup a Super Node for every local network (acts as a router)
    a) Super Node maintains an ordered list of other known super nodes in the local network group
    b) Super Node periodically selects a random list of known super nodes and checks its availability by periodic ping.
    c) Exchanges the list to known super nodes
    d) Ping neighbors and purge unresponsive nodes from its list.

3) Node A searches and connects to Super Node available in the local network
4) Node B wants to use the service.
    a) Creates a search binding query
    b) Sends it over the network using the discovery service.
5) Node B sends a query to its nearest Super Node
6) Super Node hashes the query and forwards it to list of known super nodes.
7) If any super node has the details about Node A, then it sends the query to it.
8) Node A receives the search binding query and sends a reply to Node B.
9) Node B receives the search binding answer which contains Node A advertisement
    a) Node B reads the advertisement and fetches the list of available out ports for node A.
10) Connection Established.
11) After connection establishment Node B wants to download / stream a file using globus-url-copy then 11a else 14
    a) Applies and gets a certificate from certificate authority (Grid Server)
12) Extracts the IP address available in Node A advertisement.
13) Checks the certificate of Node A
    a) Connects with Node A and sends the file using globus-url-copy and closes the connection.
14) Node B wants to download a file using mmgp-copy
15) mmgp-copy comes with two options
    a) using UDT as a transfer protocol
    b) file sent under compressed format
      a) If option (a) is selected, then
        i) Extracts the IP address available in Node A advertisement.
        ii) Connects with Node A using UDT
        iii) Sends the file and closes the connection
      b) If option (b) is selected, then
        i) Node B sends the file to compress module
        ii) File is compressed and bundled
        iii) Node B sends the compressed file to Node A
16) Node A and Node B closes the connection.

### 4.6. dWave: Video Compression/Decompression Module

Video compression has gained a lot of popularity over the past decade, especially with the advent of the internet. Given that the internet has made multimedia access pervasive, new and more efficient techniques are being developed constantly to ensure maximum compression, and as a consequence, minimum network bandwidth usage. Though many techniques have been proposed in the past regarding video compression techniques, many of them suffer from shortcomings. We propose dWave, a new video compression scheme that offers better compression and PSNR values as compared to current techniques. By utilizing the power of discrete wavelet transforms we have implemented and evaluated the algorithm that attempts to minimize the redundancies present in a video. Having implemented the compression technique, we compared the compression percentage, file size and PSNR values with existing techniques. The experimental results show a considerable improvement in performance and efficiency.

    Figure 10 shows the architectural diagram for the proposed dWave video compression algorithm. The algorithm initially obtains a video grab and split this data into the audio and video streams. The video track is initially split into its constituent frames using a third party Java's library that aids video manipulation. Once the constituent frames are obtained, and they are passed to a temporal decomposition module which aims at reducing the redundancies in between frames. For temporal decomposition, we evaluated different approaches. First approach was a simple averaging scheme (Santos Paulo J. et.al., 1995) using which multiple frames are combined into a single frame by taking an average of the pixel values. This approach, though effective from the point of view of compression ratio, has defects. The main problem was the fact that retrieving the original frames was difficult given the fact that averaging wipes out individual frame data. Similar problems are faced by other techniques such as an exponential frame combining approach and the 'ghosting' approach (Santos Paulo J. et.al., 1995). Second approach was an approach where the higher valued pixels from adjacent frames were retained and others were discarded (Brunello D et.al. 2003). Given that the constituent frames need to be recovered combining the frames is not a viable option. Instead, a set of frames being passed to the temporal decomposition module needs to have a single base frame from which redundancies in following frames can be computed. A simple example of this would be to consider every second frame as a base frame and calculate difference values

between temporally related pixels between frames. This data would be much lesser in size as compared to the original data and can be exploited to achieve higher compression ratios.

This would result in a large number of zero values being computed for every second frame. Such a redundancy can be overcome by using the deflate algorithm (L.P. Deutsch, 1996). This algorithm exploits redundancies that occur due to recurrent patterns in data and offer good compression for data comprising large numbers of frequently occurring patterns. Figure 11 shows the images after applying temporal and spatial decomposition.

Once the temporal decomposition phase is done, the frames are passed to the spatial decomposition module which removes the redundancies present within a given frame. The spatial decomposition module employs the lifting scheme which results in reduced computational complexity (Das A.et.al. 2010) (Wim Sweldens, 1998). The filter used is Cohen-Daubechies-Feauveau 9/3 filter, and the coefficients used are:

High-pass: $\{3,6,-16,-38,90,-38,-16,6,3\}*2^{(-13/2)}$
Low-pass: $\{32,64,32\}*2^{(-13/2)}$

For compressing the data, a modified version of conventional Daubuchies approach to redundancy reduction is adopted. One of the biggest problems related to discrete wavelet transform based techniques is that of the memory requirement. Discrete wavelet based techniques require a large amount of memory, and several hardware approaches have been designed to support this. However, we employ a scheme known as the lifting scheme. Apart from reducing the memory requirements, the lifting scheme also brings along with it other advantages such as easier construction of inverse wavelets and reduced computational complexity. Using such a scheme in a software implementation is essential as without this, computational time would be enormous. However, keeping in mind the end user's system which is usually equipped with such specific hardware architectures, the lifting scheme turns out to be an essential component of the proposed technique.

Preliminary tests are conducted using dWave algorithm and shows that the compression percentage, and the PSNR values obtained are significantly better when compared to existing wavelet based compression techniques. We compared the results with some of the results from existing techniques in order to get a quantitative standing of proposed dWave algorithm. The preliminary results are very promising and provide further support to our claim. The experimental test results are discussed in section 5.2.

*4.7 Quality of Service*

Network supporting deliver of multimedia and applications have to meet diverse QoS requirements. To ensure that multimedia applications render guaranteed and required QoS, it is not enough to merely commit resources. It is important to ensure end-to-end QoS of media streams, in both the networks and the end terminals. Degradation in QoS is often unavoidable. Hence a real time monitoring is done for checking level of QoS and by taking appropriate actions at the time of degradation. Both wired, and wireless networks have QoS characteristics with a different degree of variability in parameters such as bandwidth, delay and jitter. In the proposed protocol, following parameters are considered i) video clarity by comparing PSNR values, ii) file size and iii) compression percentage.

**5. Experiments and Results**
*5.1. Experimental Results of MMGP*
The proposed application layer protocol is tested, and the results are tabulated by comparing various transfers protocols used in grid computing environment. The results are compared with in-use protocol such as globus-url-copy and globus-url-copy with UDT options. GridFTP is not a transport protocol by itself. It is a high performance, secure, reliable data transfer protocol optimized for high bandwidth wide area networks. UDT is a high performance data transfer.   And, it is an alternative data transfer protocol when TCP does not work well. One of the most common cases and the original motivation of UDT is to overcome TCP's inefficiency in high bandwidth delay product networks. UDT is connection oriented unicast, duplex, and it supports both reliable data streaming and partial reliable messaging. The details about the two protocols are discussed in (J. Suresh, A. Srinivasan, A. Damodaram, 2010) and authors studied and tabulated the results for these two protocols.
Table 1 display the experimental results obtained for the proposed protocol compared with those other in-use protocols. It can be seen from the table that for the transfer of 400MB size file, globus-url-copy took 432.85 seconds, whereas mmgp-copy with UDT took 302.54 seconds. globus-url-copy also has an option of using UDT. It took 315.91 seconds which is higher than mmgp-copy did. The reason for decrease in time is due to the automatic connectivity between two clients in case of mmgp-copy, whereas in globus-url-copy the connectivity has to be initiated by the clients manually. When the option compress is used the file size is reduced and results in decreased transmission times, which come to 110.56 seconds from 302.54. The details of the option compress and its experimental results are discussed in section 5.2.

Figure 12 shows the plotted results with video size as X − axis and transfer time as Y − axis. It plots the results obtained by comparing the in-use protocols such as globus-url-copy, globus-url-copy with udt, mmgp-copy with udt and mmgp-copy with compress options.

From the plot, a drastic reduction in transfer time for mmgp-copy with compress when compared with globus-url-copy can be seen. This reduction in time is due to compressing of files. When adopting UDT option, there is slight or near equal transfer time for globus-url-copy and mmgp-copy. The slight reduction of time for mmgp-copy is due to automatic connection establishment, where as in globus-url-copy it is manual.

### 5.2. *Experimental results of video compression module (dWave)*

The proposed compression algorithm dWave was compared with other variants of transforms such as wavelet and cosine. In wavelet transforms, Haar and Daubuchies are considered and in case of cosine transform, 2D-DCT and 3D-DCT are taken into account. dWave's efficiency is checked by using these parameters i) file size, ii) compression percentage, iii) peak-to-noise signal ratio (PNSR).

*File Size*

Sample multimedia files of different sizes have been taken for compression. Table 2 tabulate and compare the results obtained from different sized files and various other compression algorithms. There is a large size difference between the original video and the compressed video. The video size has been reduced drastically. When compared to other compression algorithms, the proposed algorithm is most efficient.

A drastic reduction of size in video files can be noticed from the table. For e.g. consider the 400MB video file is reduced to 8.65MB in the proposed algorithm which is half of the 3D-DCT variant. Proposed algorithm works well when the video sizes are more, and it gives near results with other variants for smaller sizes. For example, smaller file sizes e.g.25MB, proposed algorithm reduces it to 0.52MB and 1.03MB, which nearly equal to other variants, i.e. 0.56 for Daubechies and 0.97 for Haar.

Figure 13 shows the graphical comparison of video files before and after compression. With video size as X − axis and compressed video size as Y − axis, for the proposed and adopted algorithm, there is a gradual increase in reduction of file sizes when compared with other variants.

*Compression Percentage*

Compression percentage is the ratio of difference between the original image and compressed image. This percentage gives how much quantity the original image is compressed. Compression percentage is calculated by using the following formula, CP = (Osize – Csize) / Osize  where Osize is the size of original video and Csize is the size of compressed video. Table 3 shows the obtained compression percentage for different video sizes and for various compression algorithms including the proposed one.

It can be noticed from the table that there is slight variation in percentage when compared with that of the other algorithm. However, the proposed algorithm outperforms the others, and also it obtains more compression percentage than others. This is because of using spatio-temporal combination and wavelet transforms, adopting the power of both. Consider 400MB video file in case of cosine transform variants, compression percentage obtained are 93.56% and 94.04% respectively, whereas in wavelet transforms they are 95.08% and 96.11% respectively. In case of proposed algorithm, it is 97.84%. Figure 14 illustrates the graph of obtained compression percentage for various video sizes vs. different compression algorithms. Proposed algorithm reaches a maximum of 97.84% and 97.92% compression percentage for 400 MB and 25 MB sized video files, denoting that the algorithm compresses well than the other algorithms which have a maximum of 93% to 94% compression percentage.

*PSNR Calculation*

Peak-signal-to-noise-ratio is used as a quality parameter for reconstruction of compression images or videos. Here signal is in the original data, and the noise is in the compressed data. Calculating PSNR values is used as estimation to human awareness for reconstructing quality of compressed data, i.e. higher PSNR, high quality of video.

PSNR Calculation:

Step 1: Calculate Mean Square Error [MSE]

$$d(f(x,y), f`(x,y)) = ||(f(x,y) - f`(x,y)||^2$$
$$= \frac{1}{mn}\sum_{i=0}^{m-1}\sum_{j}^{n-1}\left((f(i,j) - f`(i,j)^2\right),$$

where $f(x,y)$ and $f`(x,y)$ are original and reconstructed images respectively, m and n are image size.

Step 2:  $PSNR = 10\log_{10} 1/MSE$

Table 4 shows PSNR values obtained for different comparison techniques. Figure 15 displays the graph plotted with various bit rates vs. obtained PSNR values for a video sequence. We can notice that the desperate increase in PSNR values which are in acceptable range and efficient when compared to other algorithms.

The proposed application layer protocol was tested by using these benchmarks, i) connection establishment, ii) file size vs. download time, iii) simultaneous connections, iv) bandwidth utilization, v) flows, vi) congestion control, vii) traffic type, viii) security, ix)flow control, x) quality of service and xi) CPU usage. Table 5 shows the benchmark results. When the peer starts the service, connection establishment is done automatically in the proposed protocol and the time taken to establish the connection is reduced when compared with those other in-use protocols which has manual connection establishment. Downloading or streaming heavy sized file's matters in case of globus-url-copy and UDT protocols. Whereas in mmgp-copy, the option compress comes into rescue to reduce the file size by compressing it. Hence, transmission time is reduced. As the proposed protocol incorporates JXTA architecture as a base layer, this benchmark makes mmgp-copy a unique when compared with the other in-use protocols in grid environment. Bandwidth utilization depends on a number of connections i.e. number of peers connected in the network which indirectly employs in Simultaneous connections. Bandwidth utilization increases in mmgp-copy because of reduction of file size i.e. using compress option. CPU usage is also minimized and compensated. As the connection establishment is automatic, CPU usage time is reduced and when compress option is used CPU usage time is increased. When deploying grid concepts in peer-to-peer networks, scalability increases as there is no administrative node in wireless network concept.

## 6. Conclusion

Computational capacity is growing fast as more and more home equipments include microchips. With the advent of 3G and 4G, wireless communications are becoming more general. They offer a way for all digital equipment to interact in a more effective and faster way. As one of the JXTA design principle is ubiquity, it should be used with any kind of devices. Proposed protocol would then offer a way to harness all these wireless digital equipment and build "Wireless-Grids". Presently, this concept is specified only for delivery of multimedia and its services in grid environment in wireless mode. In future, MMGP can be made as de-facto standard for building "Wireless-Home-Grids" which would improve the transfer speed and thus increasing computational capacity without investments in new hardware. The proposed next step for MMGP concept is to investigate possibilities of integrating it with the other existing grid-middleware such as GridBus, GridSim etc. and to provide maximum functionalities of grid computing so that grid nodes can be created and deployed dynamically in wireless networks.

**Srinivasan Arulanandam** completed his ME, PhD in computer Science and Engineering at Madras Institute of Technology, Anna University, Chennai. He has finished his Post Doctorate at Nan yang Technological University, Singapore. He has 20 years of Teaching and Research Experience in Computer Science and Engineering field and one year of Industrial Experience. He has successfully guided 32 M.E projects and currently guiding 8 PhD students. He has published more than 52 Research publications in National and International journals and conferences. He is on the editorial board in Journal of Computer Science and Information Technology [JCSIT] and Review Board Member to ten reputed International Journals in Computer Science and Engineering field. Currently he is working as Senior Professor and Head, Department of Information Technology, Misrimal Navajee Munoth Jain Engineering College, Anna University, Chennai, India. His fields of interests are Digital Image processing, Face Recognition and Distributed Systems.

**Suresh Jaganathan** received his Bachelors Degree from Mepco Schlenk Engineering College, Sivakasi and M.E (Software Engg.) from Anna University, Chennai. Currently he is pursuing his PhD in Jawaharlal Nehru Technological University (JNTU), Hyderabad, in the field of Grid Computing. To his account he has published 17 papers in the area of Grid Computing and Peer-to-Peer Networks in reputed National & International Conferences. He has 17 years of experience in Teaching and is currently working as an Assistant Professor in Department of Computer Science & Engineering in Sri Sivasubramaniya Nadar College of Engineering, Anna University, Chennai. His research interests are Grid Computing, Distributed Computing and Neural Networks. He is a Member of IEEE and Life Member of CSI and ISTE in India.

**Damodaram Avula** joined as faculty of Computer Science & Engineering at JNTU, Hyderabad in the year 1989. In his 2 decades of dedicated service. He performed distinguished services to the University as a Professor, Head of the Department, Vice Principal, Director of UGC-Academic Staff College and Director, School of Continuing & Distance Education. He has successfully guided 6 Ph.D. and 2 MS Scholars apart from M.Tech projects. He is currently guiding 9 scholars for PhD and 1 scholar for MS. He is on the editorial board of 2 International Journals and a number of Course materials. He successfully executed an AICTE research project at a cost of 7 Lakhs. He has published 35 well researched papers in national and International journals. On the basis of his scholarly achievements and other multifarious services, he was honored with the award of DISTINGUISHED ACADAMICIAN by Pentagram Research Centre, India, in January 2010.

Table 1. Comparison of proposed protocol with in-use grid computing transfer protocols

| Video Size [MB] | Protocols | | | |
|---|---|---|---|---|
| | globus-url-copy | globus-url-copy with (UDT) | mmgp-copy with (UDT) | mmgp-copy with (Compress) |
| | [seconds] | | | |
| 25 | 27.95 | 15.87 | 10.67 | 5.65 |
| 50 | 51.18 | 35.62 | 29.78 | 13.67 |
| 100 | 113.93 | 60.31 | 67.27 | 25.78 |
| 200 | 205.66 | 125.98 | 123.56 | 37.93 |
| 400 | 432.85 | 315.91 | 302.54 | 110.56 |

Table 2. Comparing compression video file size (MB) for different compression algorithms

| Video Size | Cosine Transforms | | Wavelet Transforms | | |
|---|---|---|---|---|---|
| | 2D-DCT | 3D-DCT | Haar | Daubuchies | dWave |
| | [MB] | | | | |
| 25 | 1.81 | 1.43 | 0.97 | 0.56 | 0.52 |
| 50 | 3.12 | 2.63 | 2.22 | 1.52 | 1.03 |
| 100 | 6.23 | 5.72 | 4.67 | 3.26 | 2.26 |
| 200 | 11.98 | 11.54 | 9.78 | 7.45 | 4.54 |
| 400 | 25.76 | 23.78 | 19.67 | 15.56 | 8.65 |

Table 3. Comparison of compression percentage for different compression algorithms

| Video Size [MB] | Cosine Transforms | | Wavelet Transforms | | |
|---|---|---|---|---|---|
| | 2D-DCT | 3D-DCT | Haar | Daubuchies | dWave |
| | [%] | | | | |
| 25 | 92.76 | 94.28 | 96.12 | 97.76 | 97.92 |
| 50 | 93.76 | 94.74 | 95.56 | 96.96 | 97.94 |
| 100 | 93.77 | 94.28 | 95.33 | 96.74 | 97.74 |
| 200 | 94.01 | 94.23 | 95.11 | 96.28 | 97.73 |
| 400 | 93.56 | 94.04 | 95.08 | 96.11 | 97.84 |

Table 4. Comparing PSNR values different compression algorithms

| Bit-Rate [kbps] | Cosine Transforms | | Wavelet Transforms | | |
|---|---|---|---|---|---|
| | 2D-DCT | 3D-DCT | Haar | Daubuchies | dWave |
| | [dB] | | | | |
| 200 | 41.48 | 44.56 | 40.12 | 43.51 | 46.41 |
| 400 | 40.44 | 45.87 | 38.99 | 44.12 | 49.67 |
| 500 | 39.15 | 44.45 | 39.42 | 43.67 | 48.36 |
| 600 | 40.88 | 46.69 | 39.89 | 42.51 | 49.36 |
| 1000 | 44.54 | 48.98 | 38.74 | 44.28 | 49.19 |

Table 5. Process Benchmarks

| Benchmarks | globus-url-copy | globus-url-copy [UDT] | mmgp-copy [UDT] | mmgp-copy [Compress] |
|---|---|---|---|---|
| Connection Establishment | Manual | Manual | Automatic | Automatic |
| File Size Vs. Download Time | Time Increases as size increases | Depends on the network | Depends on the network | Time Decreases as the file is compressed |
| Lots-of-small files | Transfer time increases | Transfer Time increases | Transfer Time increases | Decreases |
| Bandwidth Utilization | Good | Good | Good | Very Good |
| Flows | Multiple | Multiple | Single | Single |
| Congestion Control | With TCP as its base using MIMD | either MIMD- MIMD (UDT) | MIMD (UDT) | MIMD (UDT) |
| Traffic type supported | CBR | CBR, VBR | CBR, VBR | CBR, VBR |
| Security | GSI | GSI | GSI | GSI |
| Flow control | Yes | Yes | Yes | Yes |
| QoS | Yes | Yes | Yes | Yes |
| CPU Usage | More | More | More | Less |

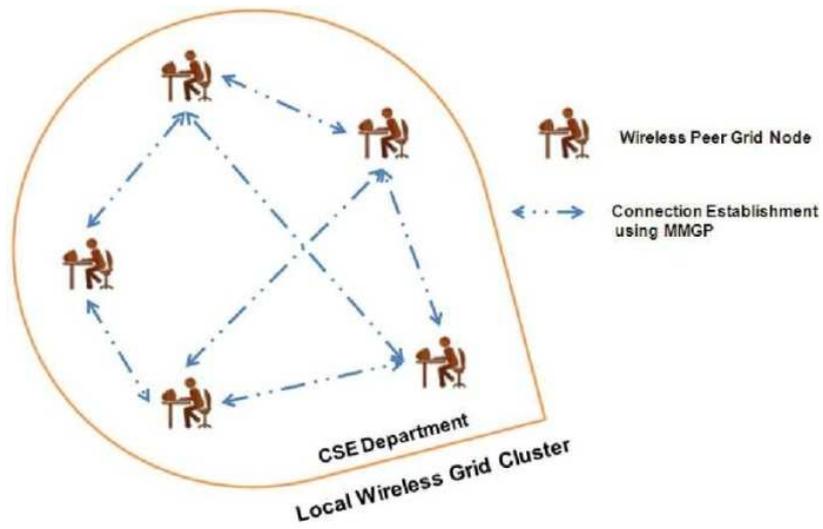

Figure 1. Local or Homogeneous Wireless Grid Cluster

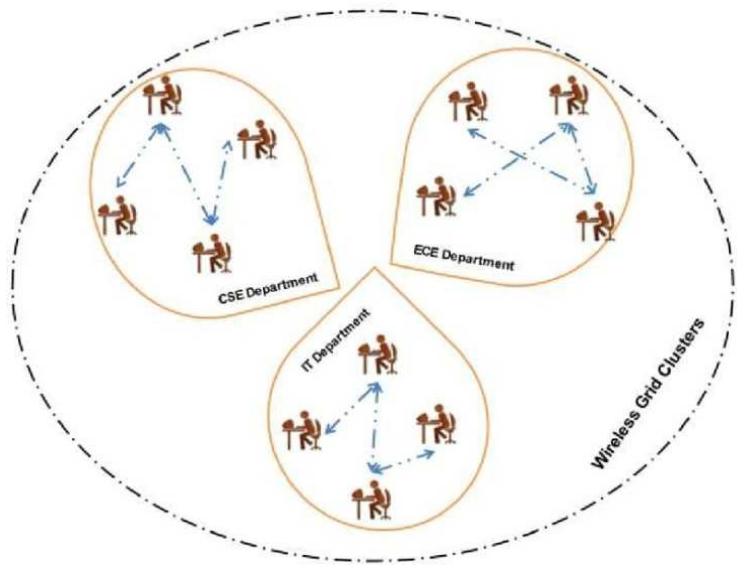

Figure 2. Wireless Intra-Grid Cluster

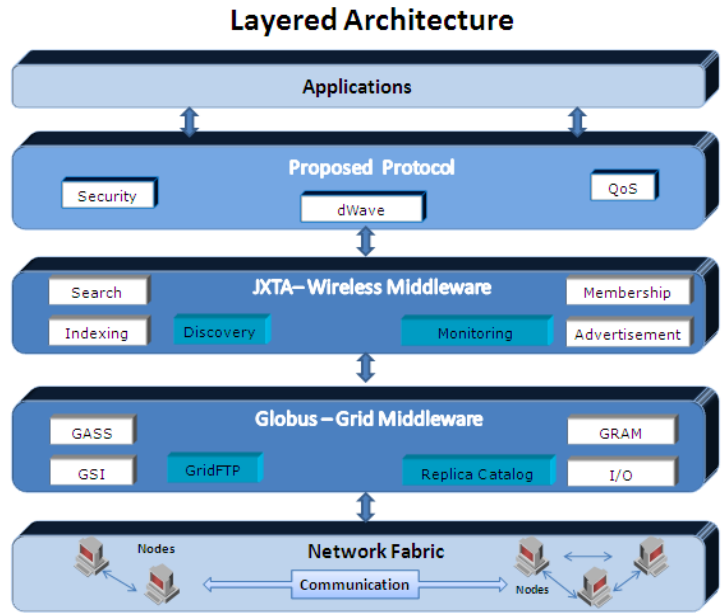

Figure   3.   Architecture of proposed MMG Protocol

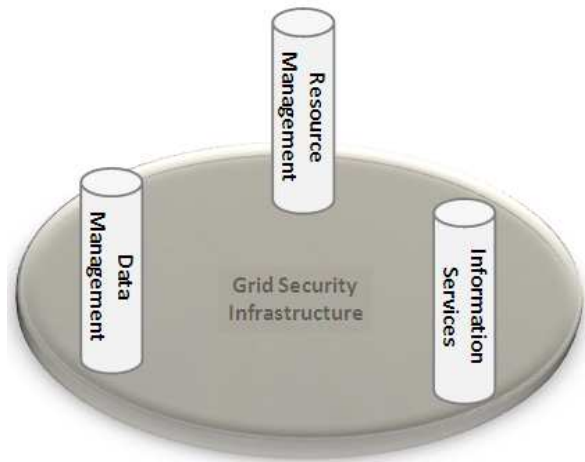

Figure 4.   Globus Toolkit Components

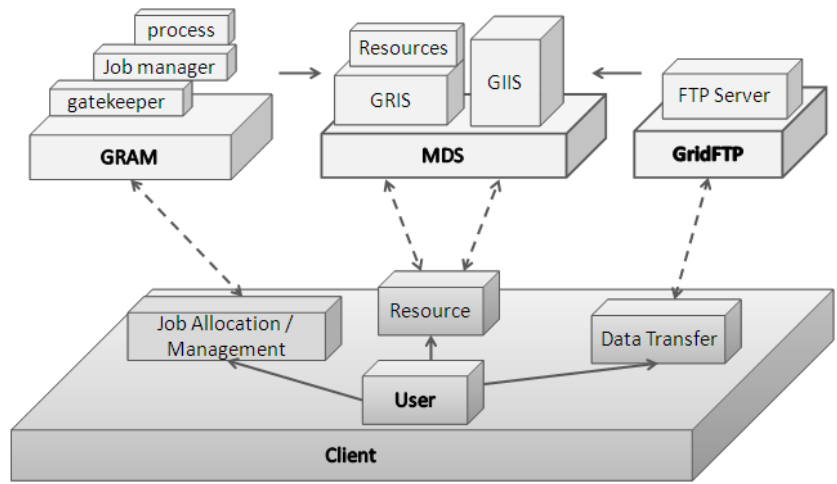

Figure 5. Globus Toolkit

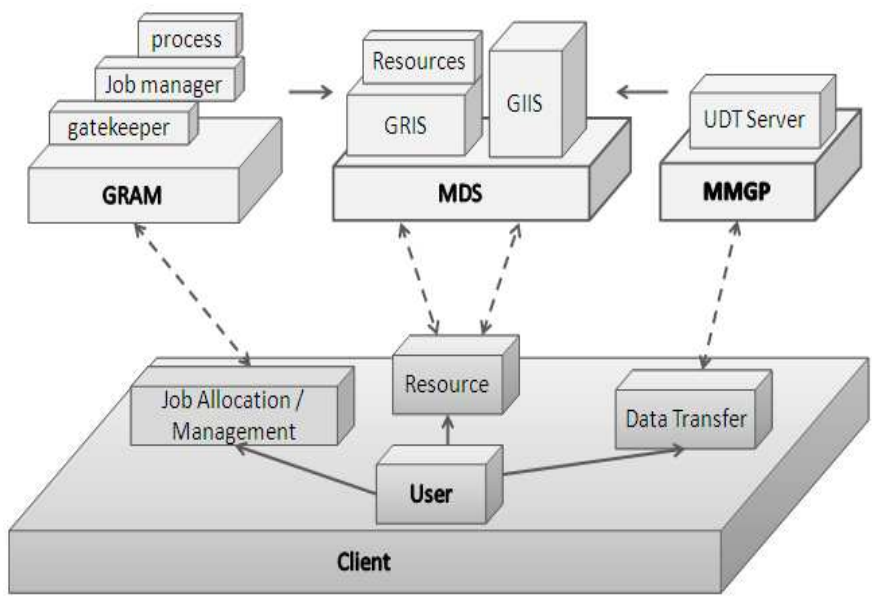

Figure 6. Incorporating MMGP in Globus Toolkit

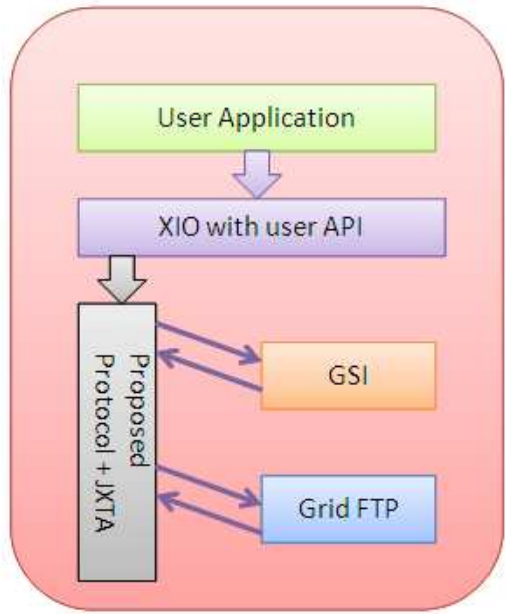

Figure 7.    Globus XIO Framework

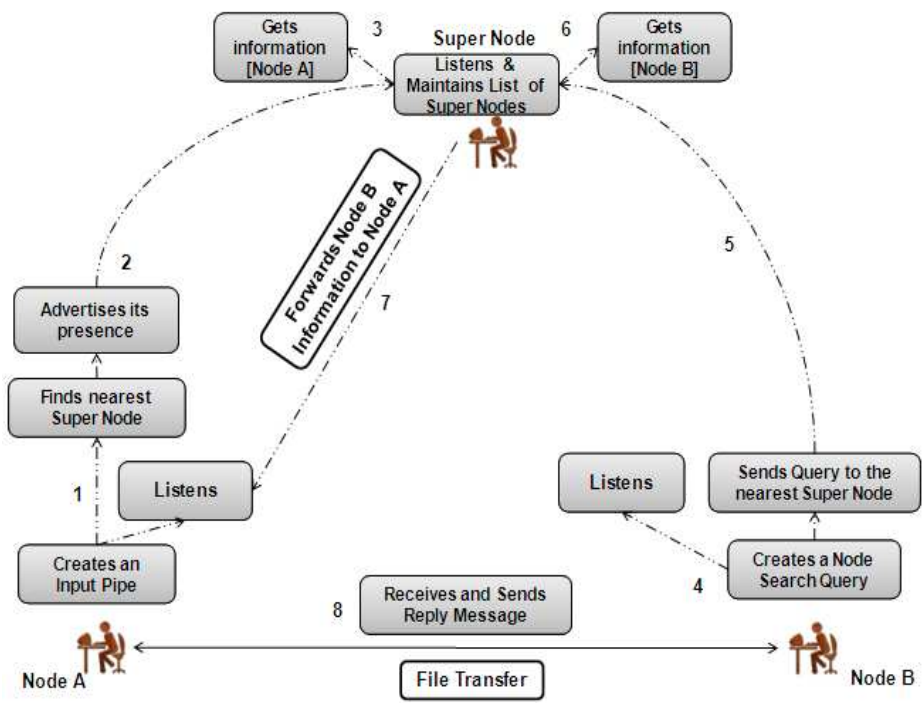

Figure   8.    Operational block diagram

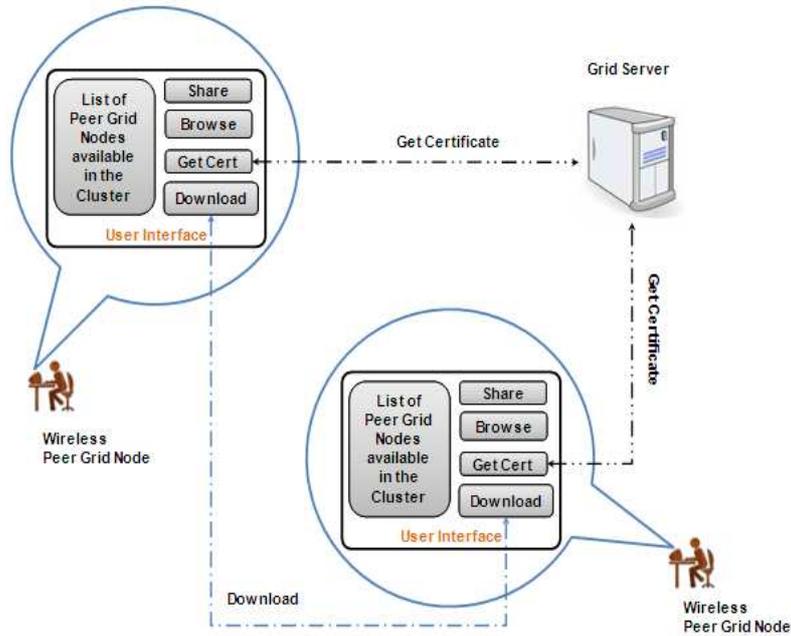

Figure 9. Working Block Diagram

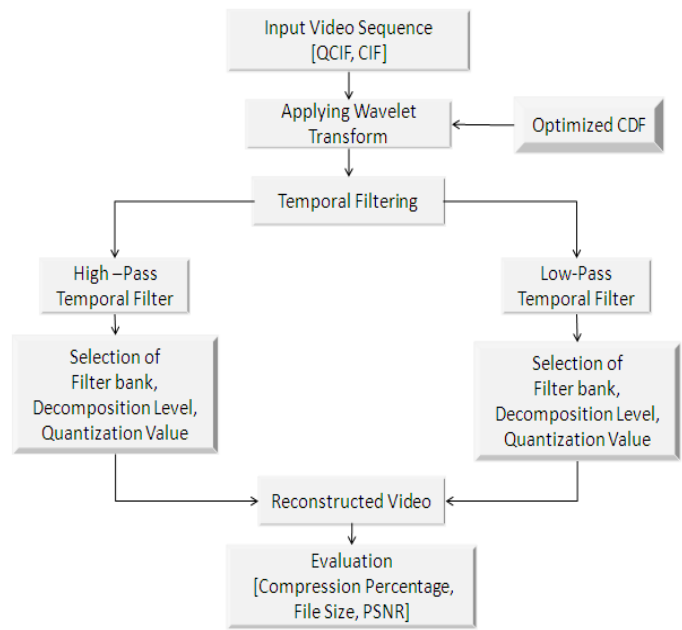

Figure 10. Proposed architecture for Video Compression [dWave]

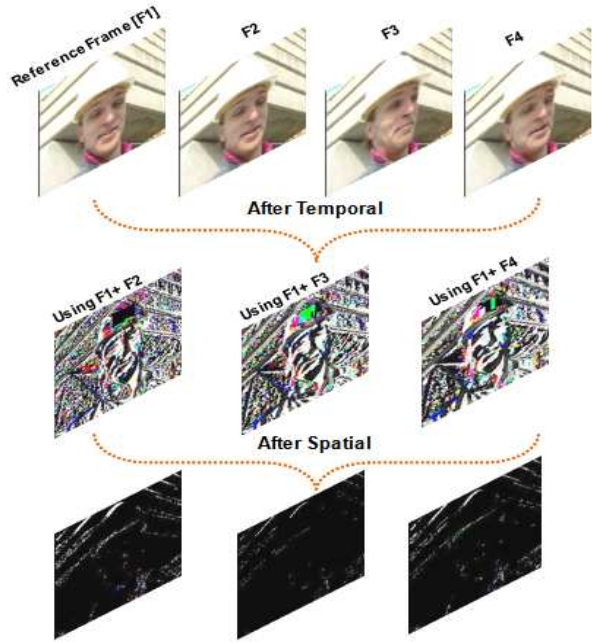

Figure 11.    Temporal-Spatial Decomposed Images

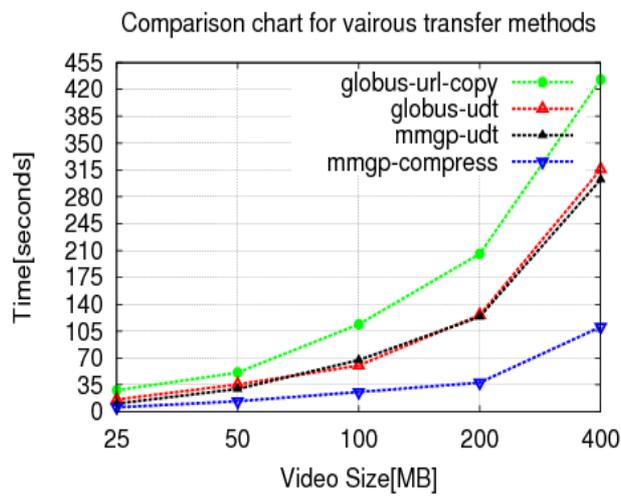

Figure 12. Comparison chart for transfer protocols in gird environment

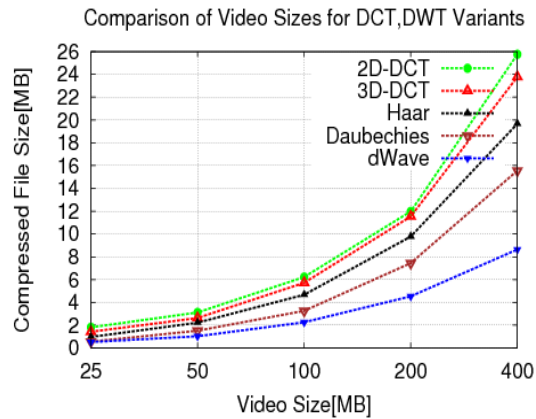

Figure 13. Comparison chart for video sizes after compression

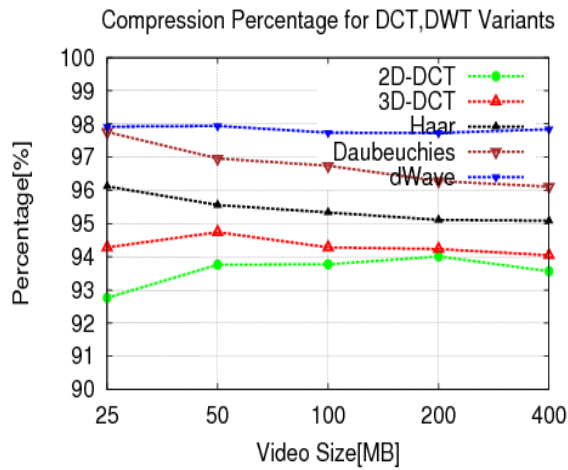

Figure 14. Comparison chart for compression percentage for various algorithms

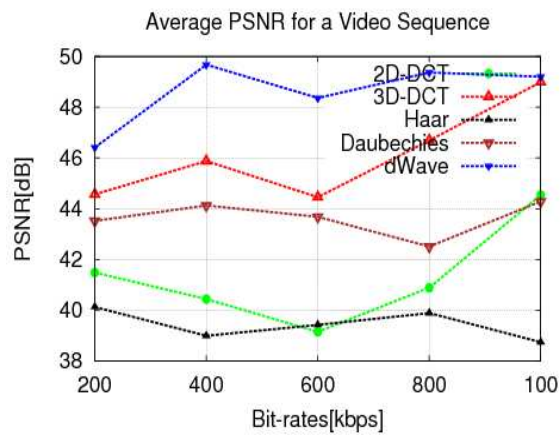

Figure 15. Average PSNR for a Video Sequence and is compared with various transforms